\documentclass{blois}

\usepackage{lineno}

\def\Journal#1#2#3#4{{#1} {\bf #2}, #3 (#4)}

\def\PLB{{\em Phys. Lett.}  B}

\def\JHEP{{\em JHEP}}
\def\EPJ{{\em Eur. Phys. J.}}
\def\JINST{{\em JINST}}

\def\be{\begin{equation}}
\def\ee{\end{equation}}
\def\bea{\begin{eqnarray}}
\def\eea{\end{eqnarray}}

\newcommand{\Photo}{}

\begin{document}
\vspace*{4cm}
\title{Searches for h(125) properties beyond the Standard Model\\ at the CMS experiment}

\author{ Teresa Lenz (on behalf of the CMS Collaboration) }

\address{DESY, Notkestr. 85, 22607 Hamburg, Germany\\ E-mail: teresa.lenz@desy.de}

\maketitle\abstracts{
The discovered Higgs boson with a mass of 125\,GeV exhibits properties which are all in agreement with Standard Model predictions. 
However, the corresponding measurements still allow for a considerable non-Standard Model behavior of the h(125).
Beyond Standard Model properties can show up in non-Standard Model decays of the h(125) or in anomalous couplings of the Higgs boson.
This article presents four different analyses done at the CMS experiment with 2012 and 2016 data that search for properties of the discovered Higgs boson beyond the Standard Model.
These include a search for lepton flavor violating Higgs decays, a search for decays of the h(125) to light Higgs bosons,
a search for anomalous contributions to the Higgs trilinear self-coupling, and finally a search for anomalous couplings of the h(125) to vector bosons.
}

\section{Introduction}

In the year 2012, a new particle has been discovered at the CMS~\cite{CMS_experiment} and ATLAS~\cite{ATLAS_detector} experiments at the LHC~\cite{ATLAS_2012_HDiscovery,CMS_2012_HDiscovery}.
This particle is found to have properties which are all in agreement with the Standard Model (SM) Higgs boson.
Despite the similarity to the SM Higgs boson, parameter fits of the Higgs boson couplings plus an additional branching fraction to beyond SM (BSM) decays reveal
that there is still a considerable space for a non-SM behavior of the Higgs boson~\cite{ATLAS_CMS_2012_Couplings}.
In Fig.~\ref{fig:First}\,(left), the best fit values for different coupling strength modifiers ($\kappa$) of the various Higgs boson couplings and a branching fraction to BSM decays are shown.
The best fit value of the BSM branching ratio is at zero but with a considerable uncertainty of 34\% at a 95\% confidence level.
These findings strongly motivate to continue searching for beyond SM properties of the h(125).
In the following, four analyses for BSM properties of the h(125) are presented.

\section{Search for lepton flavor violating decays of the Higgs boson at $\mathbf{\sqrt{s}=13\,}$TeV}

Lepton flavor violating Higgs decays are not possible in the Standard Model because of the diagonal structure of the Yukawa matrices.
This is, however, not the case in models with two or more Higgs doublets.
Therefore, an evidence of lepton flavor violation in Higgs decays would be an unambiguous proof of new physics.

Data taken in the year 2012 at the CMS experiment indicated lepton flavor violation in $h(125) \rightarrow \mu \tau$ decays with a significance of $2.4\sigma$ at 95\% confidence level~\cite{CMS_2012_LFV}.
The best fit branching ratio of $h(125) \rightarrow \mu \tau$ was estimated to a value of $0.84\pm0.38\%$.
For the same data-taking period, also ATLAS published an analysis for $h(125) \rightarrow \mu \tau$ showing a mild excess of $1\sigma$~\cite{ATLAS_2012_LFV}.
Their best fit branching ratio of $0.53\pm0.51\%$ was compatible with the one measured at CMS.
Because of the lower collected luminosity in the year 2015 at the LHC, the 2015 analyses at CMS and ATLAS did not allow a firm conclusion about the existence of lepton flavor violation in the Higgs sector.

The CMS search for lepton flavor violating Higgs decays with 2016 data~\cite{CMS_2016_LFV} analyzes two possible final states: $h(125) \rightarrow \mu \tau$ and $h(125) \rightarrow e \tau$.
All possible final states of the tau lepton are considered with the exception of $\tau \rightarrow \mu \nu_{\mu} \nu_{\tau}$ in the $\mu\tau$ final state and $\tau \rightarrow e \nu_e \nu_{\tau}$ 
in the $e\tau$ final state since these final states are overwhelmed by $Z\rightarrow\mu\mu$ and $Z\rightarrow e e$ events respectively and are therefore very hard to separate from this background.
Still, the major challenge of this search is to separate the signal events against $Z\rightarrow \tau\tau$ events since the final states of these events only differ by two additional neutrinos.
The discrimination of this background is done with the help of kinematic variables, 
like the transverse momenta of the muon or the electron which are expected to be higher for signal events since the leptons are produced promptly,
and the missing transverse energy of the event which is in general higher for background events because of the two additional neutrinos.
Furthermore, the collinear mass $M_{\mathrm{col}}$ is used which is defined as 
$M_{\mathrm{col}} = M_{\mathrm{vis}} / \sqrt{x_{\tau}^{\mathrm{vis}}}$
with $M_{\mathrm{vis}}$ being the invariant mass of the prompt lepton and the visible decay products of the tau lepton,
and $x_{\tau}^{\mathrm{vis}} = p_T^{\tau^{\mathrm{vis}}}/ \left(p_{\mathrm{T}}^{\tau^{\mathrm{vis}}} + p_{\mathrm{T}}^{\nu,\mathrm{est}}\right)$
being the fraction of the transverse momentum carried by the visible decay products of the tau lepton. 
The transverse momentum carried by the neutrino is estimated by the component of $\vec{E}_T^{\mathrm{miss}}$ in the direction of the visible tau lepton decay products. 

The event selection of this analysis includes the selection of a well identified muon or electron and a tau lepton. 
Events with additional leptons are vetoed in order to suppress background contribution from $Z$ events.
The final set of events is further subdivided into four categories which depend on the jet multiplicity and the invariant mass between two possible forward jets in order to enhance the sensitivity with respect to different production mechanisms.

The final statistical evaluation is done with a boosted decision tree (BDT).
A second approach relies on the variable $M_{\mathrm{col}}$ only.
This twofold approach is done in order to enhance the robustness of this analysis which is needed in case of a positive signal.
The BDT approach uses a wide set of variables including the transverse momenta of the final state leptons, the collinear mass, the transverse masses, 
and the spacial separation in the phi-plane between $E_T^{\mathrm{miss}}$ and the final state leptons.

No excess above the SM expectation is found, neither in the BDT nor in the $M_{\mathrm{col}}$ analysis.
Upper bounds on the branching ratio of $h(125) \rightarrow \mu \tau$ and $h(125) \rightarrow e \tau$ are set to 0.25\% and 0.61\%, respectively.
The best fit branching ratio of $h(125) \rightarrow \mu \tau$ is estimated to $0.00\pm0.12\%$.

\section{Search for light bosons in decays of the 125\,GeV Higgs boson at $\mathbf{\sqrt{s}=8\,}$TeV}

There is a variety of BSM models where more than just one Higgs boson is present.
This is for instance the case in the Next-to-Minimal supersymmetric model (NMSSM) with seven Higgs bosons.
In the NMSSM, the lightest CP-odd Higgs boson $a_1$ can be very light down to the MeV region.

Many analyses at the CMS experiment search for decays of the h(125) to light Higgs bosons.
These include searches for $h \rightarrow a_1 a_1 \rightarrow 4\mu$, $h \rightarrow a_1 a_1 \rightarrow 4\tau$, $h \rightarrow a_1 a_1 \rightarrow 2\mu2\tau$, 
and $h \rightarrow a_1 a_1 \rightarrow 2\mu2b$~\cite{CMS_2012_HToAA}.
All of these searches typically target different mass ranges of $a_1$, first because they make use of different final state particles, and second, because of the use of different analysis
and reconstruction techniques which are best suited in different phase spaces.
In this article the search for $h \rightarrow a_1 a_1$ in the four tau lepton final state is presented in a greater detail.

The search for $h \rightarrow a_1 a_1 \rightarrow 4\tau$ targets mass ranges of $a_1$ between $5-15$\,GeV.
Because of the light mass of the $a_1$, this state typically experiences a large Lorentz boost which causes the decay products of the $a_1$ to overlap.
Still, in order to have a highly efficient triggering, the search requires one isolated muon.
On the other hand, to account for the overlap of the decay products, a reconstruction of a boosted $\tau_{\mu} - \tau_X$ object is required.
This object is reconstructed as follows: The reconstruction is seeded by jets with muons, where the muon is removed from the jet. 
Afterwards, the standard tau reconstruction is performed on the rest of the jet.
If this reconstruction is successful, also the reconstruction of the boosted $\tau_{\mu} - \tau_X$ object is considered successful and the event is kept.

The search for $h \rightarrow a_1 a_1 \rightarrow 4\tau$ decays does not show an excess above the Standard Model expectations analyzing 8\,TeV data.
Also the other searches for $h \rightarrow a_1 a_1$ decays do not show any deviation from SM expectations.
The full set of exclusion limits in the $m_{a_1}$ versus $\frac{\sigma_h}{\sigma_{\mathrm{SM}}} \cdot \mathcal{B}(h\rightarrow a_1 a_1)$ plane are shown in Fig.~\ref{fig:First}\,(center).

\section{Search for pair production of Higgs bosons in the two tau leptons and two bottom quarks final state at $\mathbf{\sqrt{s} = 13\,}$TeV}

Pair production of the Higgs boson is possible in the Standard Model.
However, this process is highly suppressed because of the destructive interference of the tree-level trilinear ($\lambda_{hhh}$) and the fermionic loop contributions.
Therefore, the predicted cross section for this process is only 33\,fb in the Standard Model.
This process can be enhanced in models beyond the SM due to anomalous contributions to the Higgs trilinear coupling $\lambda_{hhh}$ and the top Yukawa coupling $y_t$.
Measuring Higgs boson pair production is therefore sensitive to both couplings.

One CMS search for Higgs pair production exploits a fully fermionic final state with two tau leptons and two bottom quarks~\cite{CMS_2016_HToHHTo2Taus2Bs}.
There are various peculiarities in this search including the mass reconstruction of the h(125) using either the two bottom quarks or the two tau leptons. 
Furthermore, the discrimination of a possible signal against $t\overline{t}$ background is challenging because of the very similar final state of $t\overline{t}$ events ($2b, \, 2\tau, \, 2\nu$).
The separation from $t\overline{t}$ is done with the help of the ``stransverse mass'' $M_{\mathrm{T2}}$~\cite{MT2} which is used as final discriminating variable.
The variable $M_{\mathrm{T2}}$ has an upper bound for $t\overline{t}$ events close to the top mass whereas it can extend to much larger values for signal events.

\begin{figure}[t]
\begin{minipage}{0.24\linewidth}
\centerline{\includegraphics[width=0.99\linewidth]{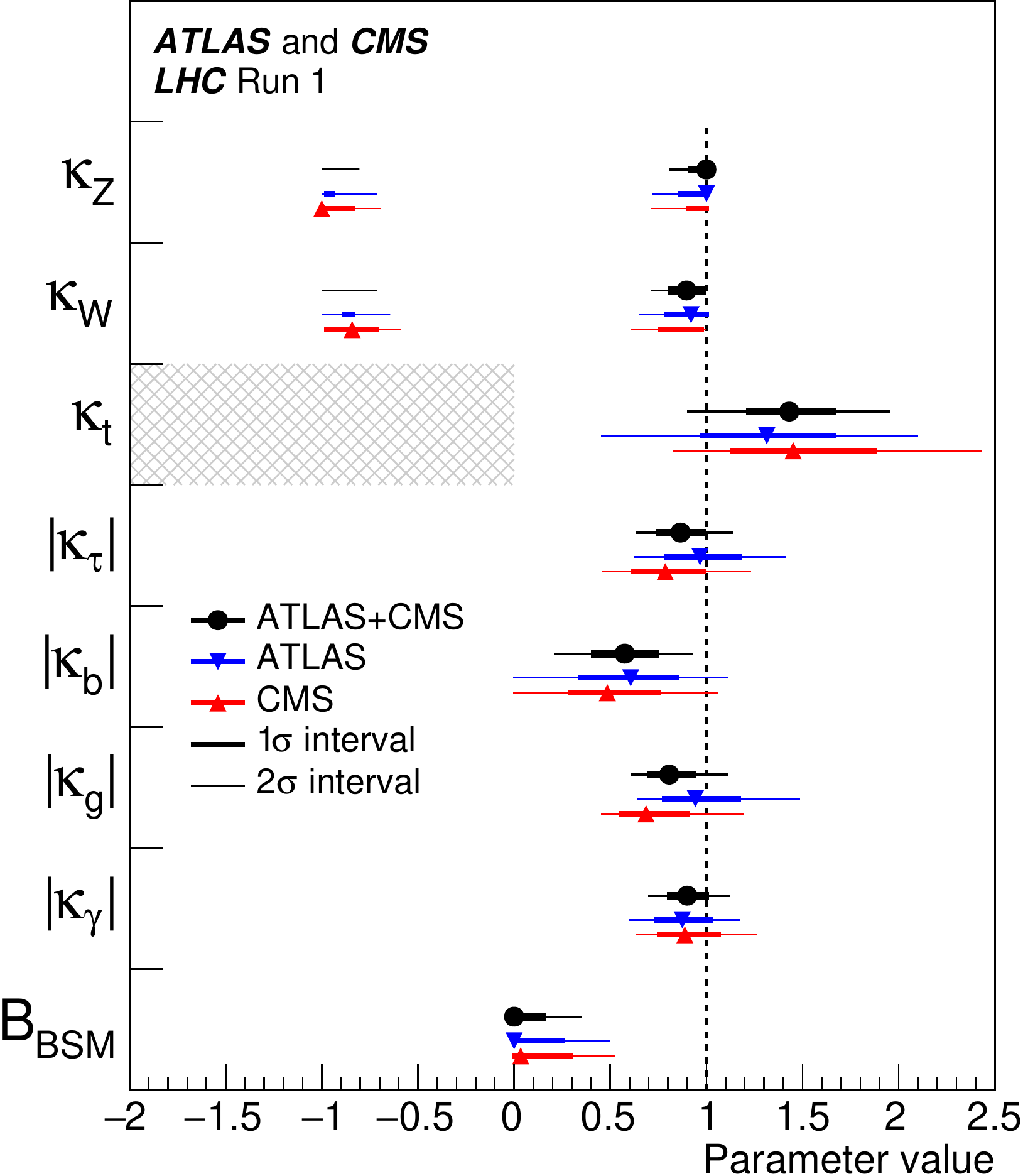}}
\end{minipage}
\hfill
\begin{minipage}{0.29\linewidth}
\centerline{\includegraphics[width=0.99\linewidth]{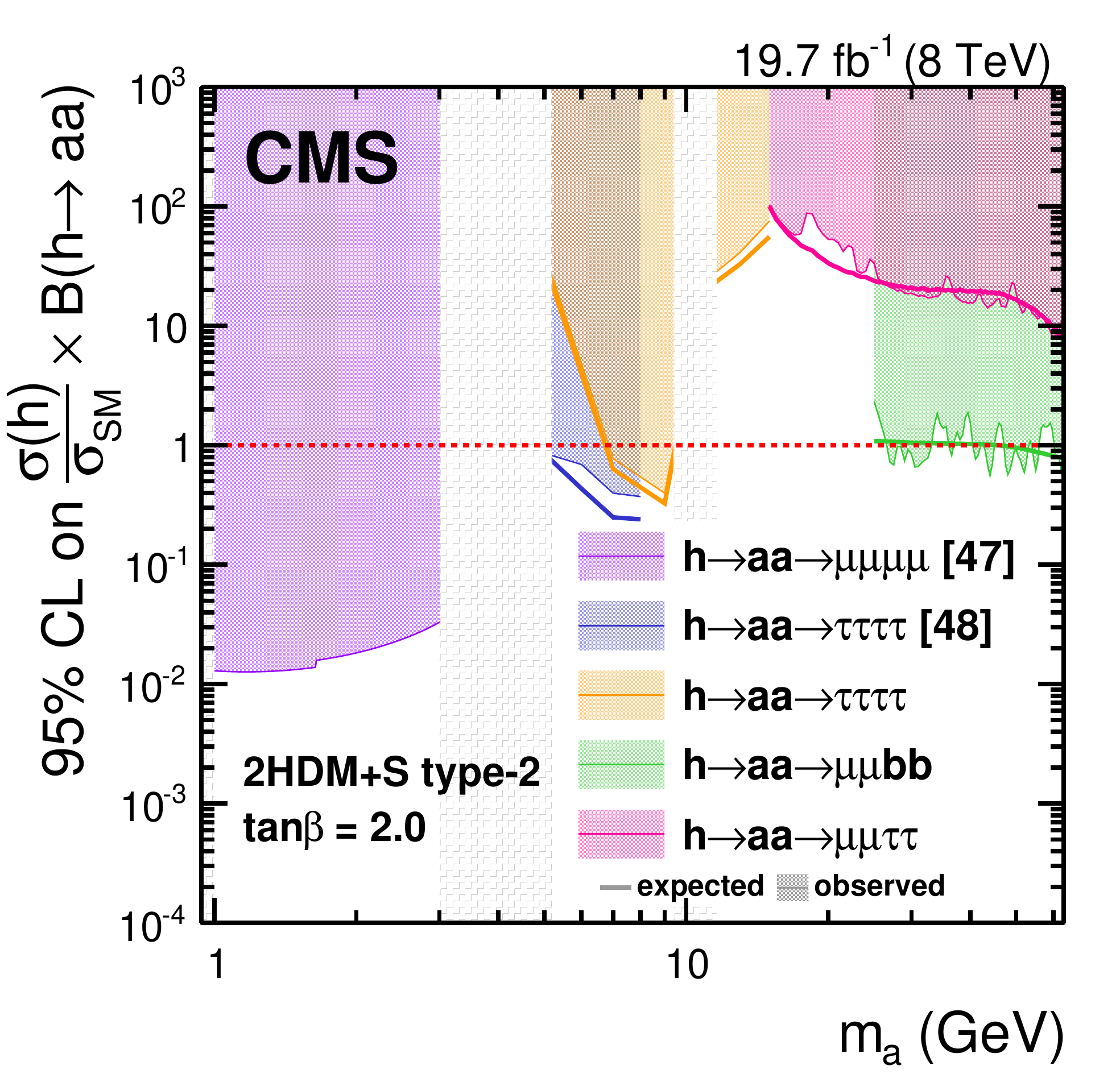}}
\end{minipage}
\hfill
\begin{minipage}{0.38\linewidth}
\centerline{\includegraphics[width=0.99\linewidth]{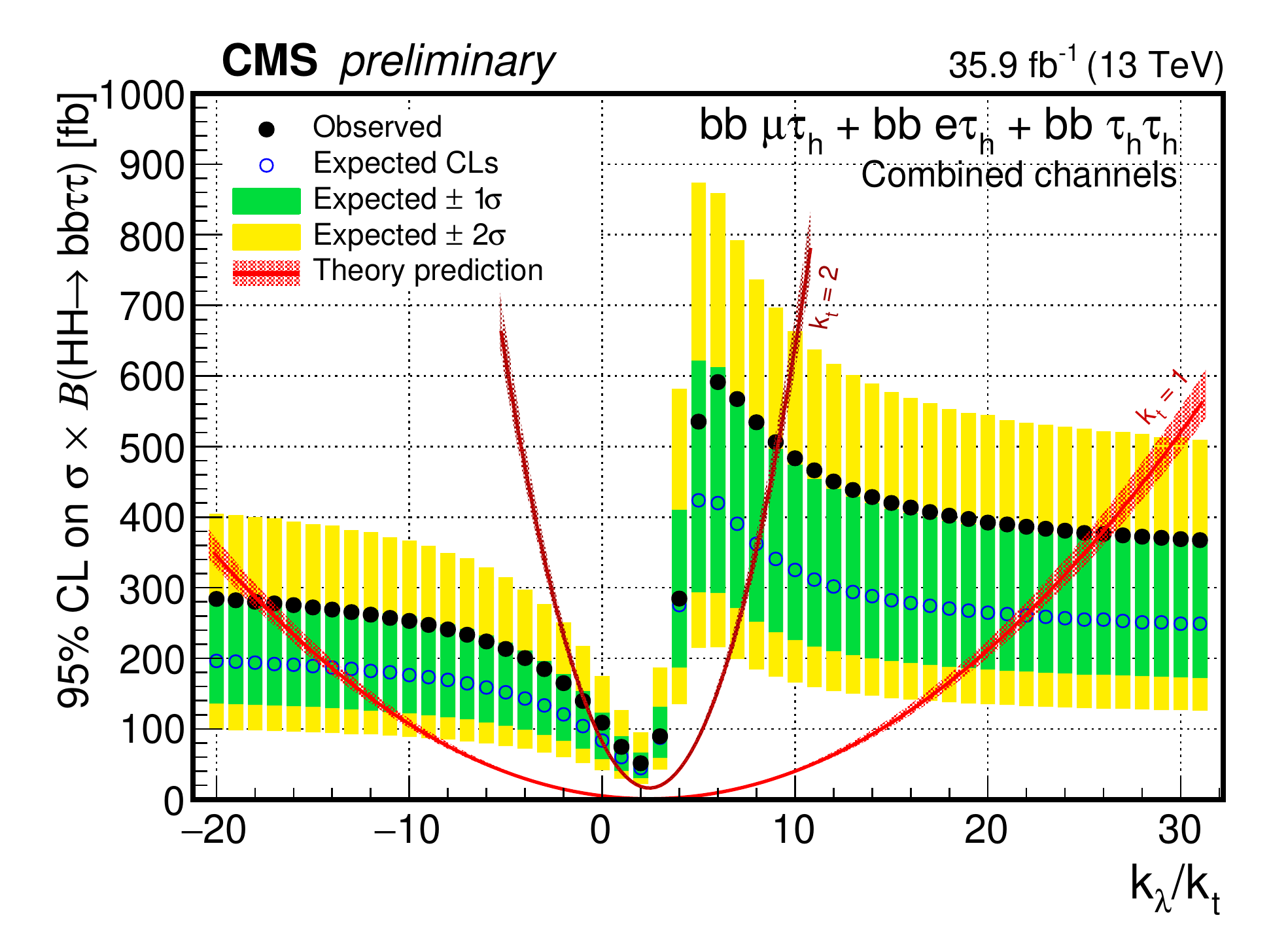}}
\end{minipage}
\caption[]{Left: Fit results for coupling strength modifiers $\kappa$ allowing a non-zero BSM branching fraction B$_{\mathrm{BSM}}$~\cite{ATLAS_CMS_2012_Couplings}. 
           Center: Expected and observed 95\% CL upper limit on cross section times branching ratio as a function of $m_{a_1}$~\cite{CMS_2012_HToAA}. \\
           Right: Expected and observed 95\% CL upper limit cross section times branching ratio as function of $k_{\lambda}/k_t$~\cite{CMS_2016_HToHHTo2Taus2Bs}.}
\label{fig:First}
\end{figure}

No excess above SM expectations is found and cross section times branching ratio upper limits are set in dependency of the ratio $k_{\lambda}/k_t$ 
with $k$ defined as the ratio of the measured coupling to the SM predicted value (see Fig.~\ref{fig:First}\,(right)).
The observed upper limit on this ratio is 28 times larger than the SM value.

\section{Constraints on anomalous Higgs boson couplings in production and decay H$\mathbf{\rightarrow4\ell}$ at $\mathbf{\sqrt{s}=13\,}$TeV}

Despite the great knowledge about the CP properties of the discovered Higgs boson, non-SM couplings to vector bosons are still possible.
The search for anomalous Higgs boson couplings to vector bosons~\cite{CMS_2016_HToVV} makes use of the four lepton final state.
Because of the higher luminosity collected in the year 2016, this search can test anomalous $hVV$ couplings for the first time not only by the decay of the Higgs boson but also by its production.
This is done by categorizing the final event yield into three categories which take into account the different production mechanisms of the Higgs boson.

This search also exploits the full decay and production kinematic information of the events by using matrix element techniques as well as up to 13 observables per event.
With matrix element techniques the set of informative observables is reduced by building up various discriminants.
These discriminants aim at separating different signal hypotheses by calculating the conditional probability $P(\vec{\Omega} | \alpha)$ 
to find the measured set of observables of an event $\vec{\Omega}$ for a given theoretical hypothesis $\alpha$.
Furthermore, the discriminants make use of the ratios of probabilities in order to reduce the systematic uncertainties.

\begin{figure}[h]
\centerline{\includegraphics[width=0.67\linewidth]{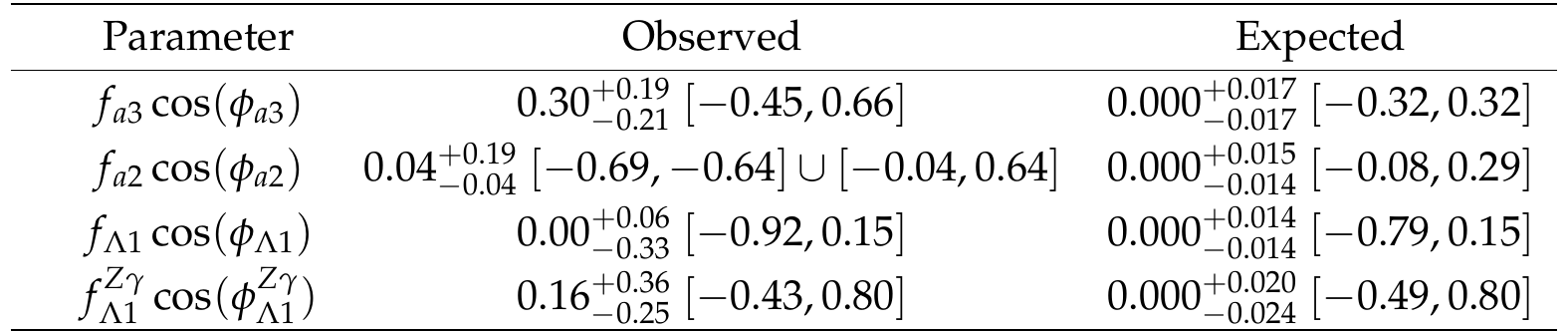}}
\caption[]{Expected and observed 68 and 95\% CL intervals on the anomalous couplings~\cite{CMS_2016_HToVV}.}
\label{fig:Second}
\end{figure}

As shown in Fig.~\ref{fig:Second}, all fit values of the anomalous couplings are compatible with Standard Model expectations within their uncertainties.

\section{Conclusion}
Higgs boson properties beyond the Standard Model can be tested with two different approaches: 
first, by searching for non-SM decays and second, by testing anomalous contributions to couplings that are also existent in the SM.
In the four here presented analyses no deviations from SM expectations are found.
Instead the excess of the 2012 analysis for lepton flavor violating Higgs decays has fully disappeared.
Furthermore, no hint for $h \rightarrow a_1 a_1$ decays is found, and no anomalous contributions to the Higgs boson self coupling or to the coupling to vector boson could be established.

\section*{Acknowledgments}
I would like to thank the CMS collaboration for the contributions to the here presented analyses as well as the CERN accelerator division for the excellent operation of the LHC.


\section*{References}
\begin{thebibliography}{99}
\bibitem{CMS_experiment} CMS Collaboration, \Journal{\JINST}{3}{S08004}{2008}.
\bibitem{ATLAS_detector} ATLAS Collaboration, \Journal{\JINST}{3}{S08003}{2008}.
\bibitem{ATLAS_2012_HDiscovery} ATLAS Collaboration, \Journal{\PLB}{716}{1-29}{2012}.
\bibitem{CMS_2012_HDiscovery} CMS Collaboration, \Journal{\PLB}{716}{30-61}{2012}.
\bibitem{ATLAS_CMS_2012_Couplings} ATLAS and CMS Collaborations, \Journal{\JHEP}{08}{045}{2016}.
\bibitem{CMS_2012_LFV} CMS Collaboration, \Journal{\PLB}{749}{337-362}{2015}.
\bibitem{ATLAS_2012_LFV} ATLAS Collaboration, \Journal{\EPJ}{C77}{no 2, 70}{2017}.
\bibitem{CMS_2015_LFV} CMS Collaboration, CMS Physics Analysis Note CMS-PAS-HIG-16-005 (2016), https://cds.cern.ch/record/2159682.
\bibitem{CMS_2016_LFV} CMS Collaboration, CMS Physics Analysis Note CMS-PAS-HIG-17-001 (2017), https://cds.cern.ch/record/2264540.
\bibitem{CMS_2012_HToAA} CMS Collaboration, arXiv:1701.02032 (2017).
\bibitem{CMS_2016_HToHHTo2Taus2Bs} CMS Collaboration, CMS Physics Analysis Note CMS-PAS-HIG-17-002 (2017), https://cds.cern.ch/record/2256096.
\bibitem{MT2} C.G. Lester and D.J. Summers, \Journal{\PLB}{463}{99-103}{1999}.
\bibitem{CMS_2016_HToVV} CMS Collaboration, CMS Physics Analysis Note CMS-PAS-HIG-17-011 (2017), https://cds.cern.ch/record/2256097.
\end{thebibliography}
\end{document}